\documentclass[fleqn,12pt,twoside]{article}

\usepackage{graphicx}

\newcommand{\AmS}{{\protect\the\textfont2
  A\kern-.1667em\lower.5ex\hbox{M}\kern-.125emS}}

\title{ Calibration and Interpixel Capacitance of a H2RG(2Kx2K) Near-IR Detector  }

\author{G. Smadja$^1$, C. Cerna$^2$ and A. Ealet$^2$\\
\\
$^1$\normalsize{IPNL, Institut de Physique Nucléaire de Lyon}\\
\small{4, rue Enrico Fermi, 69622, Villeurbanne cedex,France}\\
$^2$\normalsize{CPPM, Centre de Physique des Particules de Marseille}\\
\small{163 avenue de Luminy Case 902, 13288 marseille Cedex ,France}}
       
\begin{document}
\maketitle 
\begin{abstract}
A temporal analysis of the noise is performed, and non linearities 
are taken into account. We then extend the correlation method to groups of several pixels 
to derive the interpixel capacitance of a detector,
found to be x = -0.0263 $\pm$ 0.0020 (stat)$\pm$0.0040 (syst.)
All measurements are consistent to a sub-percent accuracy. 

\vspace{1pc}
\end{abstract}

\section{The Apparatus}
The measurements described in this paper  were carried out in a dedicated setup built
to evaluate Hawaii detectors H2RG from Teledyne (ex Rockwell).
The detector was on loan from LBNL in view of the evaluation of its 
performance when used in a spectrograph for the JDEM project \cite{spectro:refspectro}. 
The cryostat can be operated in a range of temperature extending from 100K to 160K 
with fluctuations smaller than 0.1 K, and its equilibrium temperature  
in the absence of heating is about 110K.
It was designed so as to ensure a variation rate of temperature smaller than
0.5 K/minute whatever the liquid Nitrogen flow. 
A mirror located in the cryostat ensures a uniformity of illumination 
better than 1\% over the full detector in the tests described in this paper.  
 
The readout cards were adapted  from the acquisition system of the OPERA neutrino 
experiment. Their schematic layout is shown in Figure 1, and their configuration 
on the cryostat in Figure 2. This acquisition system was actually used in the
tests performed on the SNAP Spectrograph demonstrator in the Near infrared range \cite{spectro:refsetup,spectro:refcalib}.

\section{The calibration scheme}
We adopt the standard method of calibration as in \cite{gertNIM:Finger,gertSPIE:Finger} taking advantage 
of the stochastic Poisson fluctuations from frame to frame under illumination by a Light Emitting Diode. 
The overall variance will be the sum of the contributions from readout noise, common mode noise, 
and stochastic  noise, the latter  being easy to extract since it is the only one to depend on 
illumination. The H2RG detectors have been extensively tested in the SNAP
context as reported in \cite{NIRsensors:Tarle,NIRsensors:Smith}. 
These authors also
investigated the {\it spatial} correlations of the signal noise. 
The method proposed here is a variant where the emphasis is 
on a {\it redundant} determination
of the interpixel capacitance using groups of
pixels, and a {\it temporal} analysis of the LED signal is performed, as
would occur in an actual flux measurement. Non linearities are taken into 
account.

The ADC responds to a voltage change between two frames according to  
$\Delta A = k \Delta V$  where the calibration coefficient $k$ is the product of the 
emitter-follower factor  (~0.85) and of the ADC conversion factor 
(70 $\mu V/9.5/ADCU)$, and $V$ is the pixel voltage.\\

For each LED setting, 7 to 10 consecutive measurements were performed. The first two were ignored
as the detector is not in a stationary regime, and the last five were averaged to obtain our results.
The spread between consecutive exposures with the same LED intensity was used to estimate the
measurement errors: their origin is {\it NOT} statistical, {\it NOR} due to a slow variation of
the LED intensity. 
 
\noindent
In a single pixel detector, the voltage change is linearly related 
(in the appropriate conditions) to the number of electron stored as
$\Delta V_{pix} = q\Delta N_e/ C_0$ where $\Delta N_e$ is the number of electrons stored
in the pixel, $q$ the electron charge, and $C_0$ the capacitance of a single pixel,
quoted by Teledyne as 40ff. The conversion factor $f_e = e/ADCU$ is then expected to be 
$$f_e = \frac{\Delta N_e}{\Delta A} = \frac{C_0}{kq} \sim 2.15$$
The stochastic fluctuation of the charge on the pixel is 
$\delta Q = q\delta N_e$ , and the variance (time average) of the ADC readouts  between two frames 
arising from Poisson fluctuations is
$<(\delta A)^2> = (kq/C_0)^2 <\delta N_e^2> =  (kq/C_0)^2 \Delta N_e $
Where $\Delta N_e$ is the accumulated number of electrons between the two frames considered
(to be distinguished from the fluctuation $\delta N_e$ of this number).

We can substitute  $\Delta N_e = \Delta Q/q = C_0 \Delta V/q = C_0 \Delta A/(kq)$ which yields 
\begin{equation}
(\delta A)^2 =  \frac{kq}{C_0} \Delta A  = 1/f_e \Delta A 
\end{equation}
The conversion factor $f_e = e/ADCU$  is the inverse of the slope in the relation
between the variance of the stochastic contribution to the noise (in ADC units) 
in a given time, and the flux accumulated during the same time (also in ADC units).
The capacitance $C_0$ of the pixel can be obtained once $k$ has been measured. 

\noindent
In a multipixel detector, we adopt a description close to the one 
proposed by \cite{IRdet:mcCullogh}, but more general. 
The charge in pixel $i$ is now also dependent on the voltage in pixel $j$, and the 
relation is given by the electrostatic influence matrix. 
$\Delta Q_i = C_{ij} V_j$ (summation is implied). The ADC response in pixel $i$, $\Delta A_i$
will be given by:
$$\Delta A_i = kq (C)^{-1}_{ij} \Delta N_j$$
where $\Delta N_j$ is the change in the number of electrons in pixel $j$. 
The stochastic noise of pixel $i$ can again be derived from the Poisson fluctuations in the pixels

$$(\delta_i)^2 = (kq)^2 (C)^{-1}_{ij}(C)^{-1}_{il} \delta N_j  \delta N_l$$
If diffusion is negligible, there is no correlation
in the numbers $N_i$ for different pixels,and the time average of 
the product is 
$<\delta N_j  \delta N_l> = \delta_{jl} \Delta N_j$ (where $\delta_{jl}$ is the Kronecker matrix,
and $\Delta N_j$ the nb of electrons collected in pixel $j$ between consecutive frames.
Using $\Delta N_j = 1/(kq) C_{jm} \Delta A_m$  we then find

$$(\delta_i)^2  = kq \sum_j (C)^{-1}_{ij} C^{-1}_{ij}C_{jm} \Delta A_m $$
So that 
\begin{equation}
(\delta_i)^2 =  kq \sum_j  C^{-2}_{ij} C_{jm}\Delta A_m
\end{equation}
This general expression will now be simplified
by the use of convenient approximations of the matrices $C$ and $C^{-1}$.
 If we consider only the coupling of adjacent pixels, with $x = C_j/C_0$ ($x$ is negative) 
in a homogeneous detector, as in \cite{gertNIM:Finger}, the influence matrix $C$ for a group of 
5 pixels centered on pixel $i = 0$  as shown in Figure 3b) is $C = C_0 M $, where  the $(5 \times 5)$ 
matrix $M$ is of the form 
$$
M = 
\left(
\begin{array}{c c c c c }
 1 & x & x & x & x \\
 x & 1 & 0 & 0 & 0 \\
 x & 0 & 1 & 0 & 0 \\
 x & 0 & 0 & 1 & 0 \\
 x & 0 & 0 & 0 & 1 \\
\end{array}
\right)
$$
This matrix can be inverted exactly, but as the electrostatic coupling ratio $x$ will
be found to be small, the inverse matrix can be obtained by substituting $-x$ to $x$ when  
the second order terms $x^2$ are neglected.

We have now obtained for the single pixel noise of a homogeneous detector
\begin{eqnarray*}
\delta_0 &= &kq C^{-1}_{0j} \delta N_j \\
&= & k\frac{q}{C_0}(\delta N_0 - x\delta N_1 - x\delta N_2 - x\delta N_3 - x\delta N_4)
\end{eqnarray*}
evaluating the variance from both sides of this formula, we find:
$$ <(\delta_0)^2> = (\frac{kq}{C_0})^2 (\Delta N_0) (1 + 4x^2)$$
As the illumination is uniform, all fluxes are equal and the mean flux in pixel $i=0$ is
$\Delta N = \frac{C_0}{kq(1 - 4x)} \Delta A_0 = f_e \Delta A$
\begin{equation}
<(\delta_0)^2>  = \frac{kq}{C_0 (1-4x)} \Delta A_0 = s_1 \Delta A_0 
\end{equation}
Under a uniform illumination, the conversion factor (e/ADCU) was however found to be 
$$f_e = \frac{\Delta N}{\Delta A} = \frac{C_0}{kq(1 - 4x)}$$
so that 
\begin{equation}
f_e =\frac{1}{s_1(1-4x)^2} 
\end{equation}

As $x$ is negative, the conversion factor will 
now be smaller than the inverse of the slope, and smaller than expected for a 'single pixel'. 
The edge effect from adjacent pixels is expected to decrease for larger groups of pixels, 
and the relation between slope and conversion factor
should be closer to the single pixel case.
\section{Noise fluctuations for groups  of pixels}
The previous formulae allow us to measure the ratio $x = C_i/C_0$ by comparing the relation
between noise and flux for groups of pixels, as this changes the weight of the contribution 
of the neighbouring pixels.\\

\noindent
{\it \bf 2 pixel groups (as shown in Figure 3a)}\\

The $8\times 8$ capacitance matrix $C$ to be considered follows from Fig. 3 a):
\vskip 0.5 true cm
$$ C = \left( \begin{array}{c c c c c c c c}
             1 & x & 0 & 0 & 0 & x & x & x \\
             x & 1 & x & x & x & 0 & 0 & 0 \\
             0 & x & 1 & 0 & 0 & 0 & 0 & x \\
             0 & x & 0 & 1 & 0 & 0 & 0 & 0 \\
             0 & x & 0 & 0 & 1 & x & 0 & 0 \\
             x & 0 & 0 & 0 & x & 1 & 0 & 0 \\
             x & 0 & 0 & 0 & 0 & 0 & 1 & 0\\
             x & 0 & x & 0 & 0 & 0 & 0 & 1 \\
            \end{array}
     \right) 
$$
\vskip 0.5 true cm   

For a typical value of $x = -0.02$, the diagonal elements of the inverse
matrix vary from 1.0004 to 1.0016, while the off-diagonal elements of adjacent 
pixels differ from  0.02 by less than $5 10^{-5}$, the other coefficients 
are smaller than $4 10^{-4}$. Neglecting all small offsets, we find:\\ 
\begin{eqnarray*} 
\delta_2 = &\frac{1}{C_0}( \delta Q_1(1-x) + \delta Q_2(1-x) \\
             &   -x \sum_{i=3,i=8} \delta Q_i)
\end{eqnarray*}
$$(\delta_2)^2 = (\frac{k\delta Q}{C_0^2})^2 (2 - 4x + 8 x^2)\sim
                 (\frac{k q}{C_0})^2 \Delta N $$
and substituting for the expression of $\Delta N_0$ in terms of the observed ADC shift $\Delta A_0$
in 1 pixel between frames\\
$\Delta N_0 =  \frac{C_0}{kq(1-4x)}\Delta A_0$
\begin{equation} 
(\delta_2)^2  = 2 \frac{k q}{C_0 (1-4x)} (1 -2x + 4x^2)\Delta A_0
\end{equation}
\noindent
{\it \bf 5 pixel groups  (Figure 3b)   }\\ 
We find in the same way:
\begin{eqnarray*}
\delta_5 = &\frac{1}{C_0}(\delta Q_0(1-4x) + (1-x) \sum_{i=1,i=4}\delta Q_i \\
            &-2x\sum_{i=5, i=8}\delta Q_i -x\sum_{i=9,i=12}\delta Q_i)
\end{eqnarray*}
So that 
$$(\delta_5)^2 = (\frac{k\delta Q}{C_0})^2 (5 -16x + 40x^2) \sim (\frac{kq}{C_0})^2 \Delta N_0 (5-16x)$$  and using as before the relation between $\Delta N_0$ and $\Delta A_0$, the ADC shift:
\begin{equation}
(\delta_5)^2 = (\frac{kq}{C_0})\frac{5-16x+ 40x^2}{1-4x} \Delta A_0
\end{equation}
\noindent
{\it \bf 9 pixel groups (Figure 3c)}\\
and for 9 pixels  forming a $(3\times3)$ square (see Figure 3):
\begin{eqnarray*}
\delta_9 = &\frac{k}{C_0}(\delta Q_0(1-4x) + (1-3x)\sum_{i=1,7(odd)}\delta Q_i \\
           &+(1-2x)\sum_{i=2,8(even)}\delta Q_i  -x\sum_{9,20} \delta Q_i)
\end{eqnarray*}
$$
(\delta_9)^2 = (\frac{k\delta Q}{C_0})^2 (9-48x + 80x^2) \Delta N
$$ 
\begin{equation}
\delta_9^2 =  \frac{kq}{C_0} \frac{9-48x+80x^2}{1-4x} \Delta A_0
\end{equation}

\section{The measurement  method}

The cryostat temperature was set at 110K, the equilibrium value in the 
absence of any heating, during these tests.
In contrast with previous publications, which used the spatial 
correlations under different illumination conditions 
(\cite {Correlations:Brown}), we have used the time variation of the
signal to evaluate the noise and the correlation, as a training for
the flux measurements anticipated in the near future.   
The frames are grouped into 'bursts',
the number of frames in each burst decreasing  from 
60 (LED current  of 10 $\mu$A), to 15 (LED current of 100 $\mu$A). 
The variance of the differences of readouts  for consecutive frames 
is then evaluated for each pixel, and each burst, 
and the measurement is repeated  'up the ramp' until saturation 
is reached. The results are then averaged over all pixels for each burst. 
At any LED setting at least 10 exposures are taken to allow 
control of the error estimates.

\subsection{Non-linearities}

The  distribution of the flux in ADC units averaged over all bursts
for all pixels at a typical setting with a LED current of 40 $\mu$A is shown 
in Figure 4 a). It is seen that in the absence of flatfielding 
corrections, its spread  is 3.5\%. This good homogeneity of the
detector in the analysis window allows averageing of the measured
properties over all pixels (the flux and variance of each pixel is still
evaluated independently). A linear variation of $dADC/dt$ 
and of the single pixel noise along the ramp 
is then observed in Figure 4 b).  As the flux measurement is 
perfectly reproduced from one exposure to the next, 
the flux variation is an indication of 
the non-linear response of the system. The most likely source 
of this behaviour is the output FET, and the observations suggest a 
decrease of the transconductance as the grid voltage decreases with 
an increasing number of trapped electrons in the pixel well.  
The slopes of the variance and of the flux (normalised to their value 
at an output ADC value of 19000 ADCU) are respectively 
$a_{var} = 0.000388$ and $a_{flux} = -0.000336$, and they 
correspond to similar (and opposite) relative variations, 
as expected if the equivalent thermal resistor is the inverse of the 
transconductance. The effective conversion factor given
by the ratio variance/flux cannot be obtained without further 
corrections. All the following analyses have been performed 
with 2 different extrapolations, namely to the middle (32500) and to 
the lower values (10000) of the ADC dynamical range. To crosscheck the 
validity of the corrections, different LED illuminations will be compared. 
The conversion factors found are expected to
differ by 5\% as a consequence of the non linear response, but  
the interpixel capacitance obtained should be the same up to systematic
errors. 
\begin{table}[htb]
\caption{Readout Noise Horiz./Vert. pairs}
\label{table:1}
\newcommand{\m}{\hphantom{$-$}}
\newcommand{\cc}[1]{\multicolumn{1}{c}{#1}}
\renewcommand{\tabcolsep}{2pc} 
\begin{tabular*}{0.5 \textwidth}{l c}
\hline 
Pixel & var(ADCU)  \\ 
group &   \\
\hline
1           & 121.02$\pm$ 0.77   \\    
2-H         & 249.17$\pm$ 1.50   \\
2-V         & 261.24$\pm$ 1.79   \\
\hline
\end{tabular*}
\end{table}
\subsection{The 2 pixel pairs}
It is intructive to consider separately 
horizontal (readout direction) and 
vertical  pairs of pixels. Horizontal pairs are read consecutively, while 
vertical pairs are separated by the time interval needed to complete 
the readout of the intermediate pixels. The variance quoted in 
Table 1 is  obtained in the absence of any illumination by  summing 
2 pixels, measuring the difference between consecutive frames, estimating 
its variance for each burst (of typically 60 frames), averaging over all bursts, 
and then over all pairs. The result for the mean variance is given.

It is seen that intrinsic voltage changes in the chip during the readout impact the variance a level of about 4\%. Analyse is in progress to shown soon results including a temporal common mode correction by subtracting a reference channel.


\subsection{The 5 and 9  pixel configurations}

The measured values of the variance and the flux for each pixel
grouping is shown in figure 5 for 12 illumination conditions. 
The expected linear relation (after correcting for the non-linear response!)
is observed in Figure 5 for the data and the 
slopes found from the extrapolation to lower flux values are given 
in Table 2.

The ratio $x$ of the interpixel capacitance to the pixel capacitance $C_0$
 is found from the comparison of the slopes for 1,2,5,and 9 pixels:
To modify the impact of the fluctuations from adjacent pixels, we now
evaluate their contribution to the variance of larger groups of 
pixels. The interpixel capacitance $x = C_c/C_0$ can then be derived
for each pixel grouping from equations (5),(6),(7).

\subsection {Conversion factor and effective Interpixel capacitance} 
\begin{table}[!htb]
\caption{slopes}
\label{table:2}
\newcommand{\m}{\hphantom{$-$}}
\newcommand{\cc}[1]{\multicolumn{1}{c}{#1}}
\begin{tabular*}{0.5 \textwidth}{l c } 
\hline
pixel   &    Slope          \\
group   &                            \\ 
\hline
1       &    0.4021 $\pm$ 0.0026     \\        
2       &    0.8458 $\pm$ 0.0049    \\
5       &    2.2139 $\pm$ 0.0218   \\
9       &    4.2095 $\pm$ 0.0550     \\

\hline
\end{tabular*} 
\end{table} 
The values of the interpixel capacitance found in all cases are given in Table 3
with the linearity correction extrapolated to the lower ADC range (10000).
They are compatible, with small residual systematic shifts, and we shall perform a 
weighted average.
\begin{table}[htb]
\caption{The Slope ratios and Interpixel capacitance}
\label{table:3}
\newcommand{\m}{\hphantom{$-$}}
\newcommand{\cc}[1]{\multicolumn{1}{c}{#1}}
\renewcommand{\tabcolsep}{0.70pc} 
\begin{tabular*}{0.8\textwidth}{lll}
\hline
Group         & ratio  & x \\
\hline
s2/s1  &   2.1028 $\pm$ 0.0183  & -0.02455 $\pm$ 0.00416   \\
s5/s1  &   5.5063 $\pm$ 0.0651  & -0.02947 $\pm$ 0.00352   \\ 
s9/s1  &  10.4696 $\pm$ 0.1529  &  -0.02916 $\pm$ 0.0029 \\ 
\hline
\end{tabular*}
\end{table}
$$ x =  -0.0282 \pm 0.0020\qquad $$

The  value $x = -0.0282 \pm 0.0020$. The same
averages performed at ADCU = 30000 leads to $x = -0.02440 \pm 0.001$ 
For our final result, we take the average of the two estimates and attribute a 
systematic error of 0.0040 to account for the small difference 
in the non-linearity corrections (mid range and low range).

$$ x = \frac{C_{int}}{C_0} = -0.0263 \pm 0.0020 \hbox{(stat)} \pm 0.0040 
                                                 \hbox{(syst)}$$

The (single pixel) conversion factor in the lower ADC 
range can now be obtained as 
\begin{eqnarray*}
f_1 & = &\frac{\Delta Q_0}{\Delta ADC_0} =\frac{1}{s_1(1-4x)^2} \\
    & = &\frac {2.487}{(1-4x)^2} = \frac{2.487}{1.221} = 2.036(e/ADCU) 
\end{eqnarray*} and it is 10\% larger at full well.
The conversion factors found in the 2 extrapolation methods 
differ by 5\%, as expected from the nonlinear behaviour seen in figure 4b): 
they are evaluated at different values of the ADC range. The result is
very close indeed to the value of 2.15 derived from an assumed 
pixel capacitance of 40ff in section 2.     
\subsection{Diffusion and actual Interpixel capacitance}
The effective value of the interpixel capacitance found would {\it NOT} be 
the true value if 'fast' diffusion would occur as suggested by 
\cite{NIRsensors:Tarle}:
while the interpixel capacitance increases the fluctuations, diffusion from one
pixel to the adjacent ones would reduce them. 
A detailed computation shows that the contribution to $x_d$ from diffusion
is equal to the fraction $f_d$ of electrons migrating to the adjacent pixel for 2 and
5 pixel groups, but is only $x_d = f_d/3$ for the 9 pixel group. Given the consistency
of the previous results between 9 and 5 pixel groups, we obtain  $f_d <0.037$ (3 $\sigma$ limit). 
\section{Conclusions}
A general correlation method has been proposed, which allows
strong cross-checks for internal consistency of the observations
between different pixel groups. We have shown that the 
non-linearities which are seen are consistent
with the effect of a transconductance variation in the output FET of
the detector,and that they can be corrected to a sub-percent accuracy. 
The remaining systematic errors are in the $ 10^{-3}$range. It has
also been shown that the use of the reference channel have to studied in that frame of work. The imapct of that studies will be shown in a forthcoming paper. The calibration of the readout 
set-up will be studied in order to describe the noise performance 
achieved under different conditions. 

\section{Acknowledgements}
We thank all the institutions who have supported us during this work: Universit\'e Claude Bernard Lyon 1,
The IN2P3/CNRS institute, and the engineers and technicians at IPNL and CPPM who have contributed to
the apparatus JC Ianigro, A. Castera, and in particular C. Girerd who has designed the readout electronics.
We are indebted to C. Bebek (LBNL) for lending us the H2RG detector, and to G. Tarle, M. Schubnell, and 
R. Smith for many questions and suggestions.

\newpage
\begin{figure}[htb]
\vspace{9pt}
\includegraphics[width=0.9\textwidth,clip]{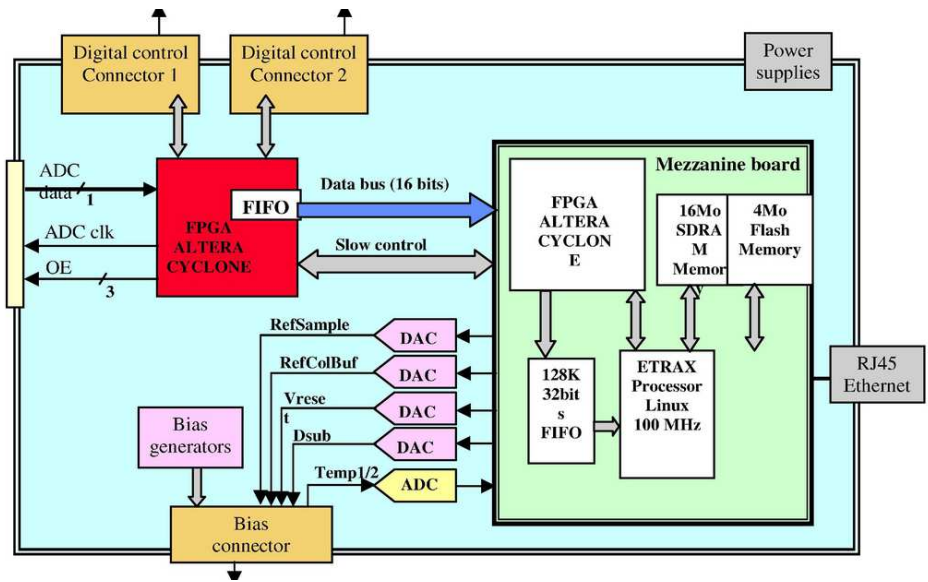}
\caption{Scheme of the acquisition card}
\end{figure}
\begin{figure}[htb]
\vspace{9pt}
\includegraphics[width=0.9\textwidth,clip]{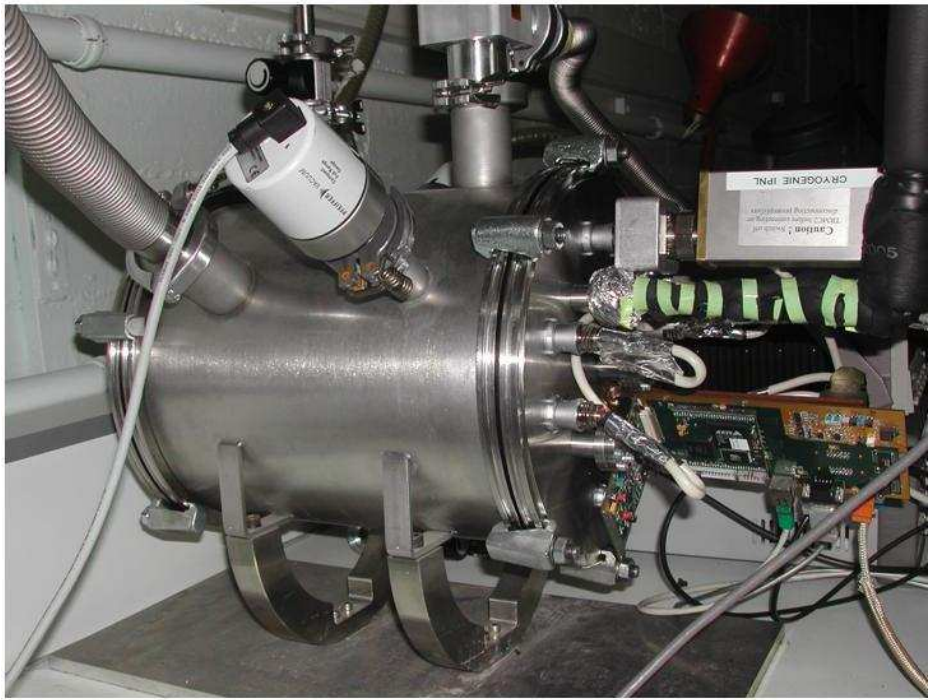}
\caption{Cryostat with the readout cards}
\end{figure}
\newpage
\begin{figure}[htb]
\begin{center}
\includegraphics[width=0.4\textwidth,clip]{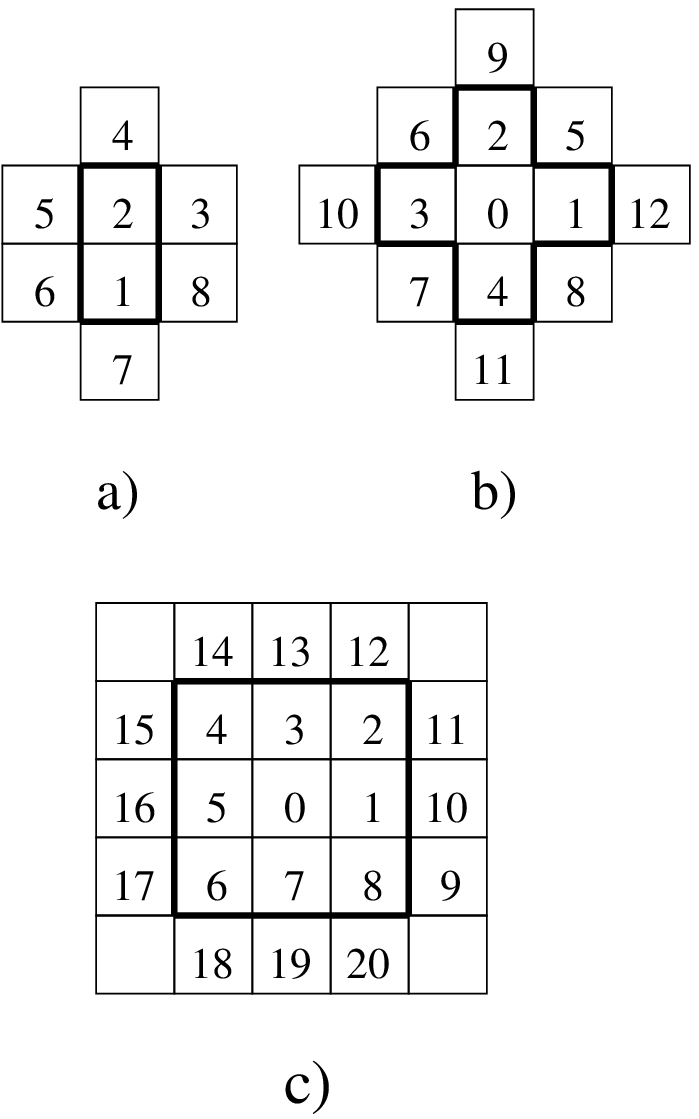}
\caption{Map of the  2,  5, and 9 pixels groups}
\end{center}
\end{figure}

\begin{figure}[htbp]
\begin{center}
\includegraphics[width= 12 cm,clip]{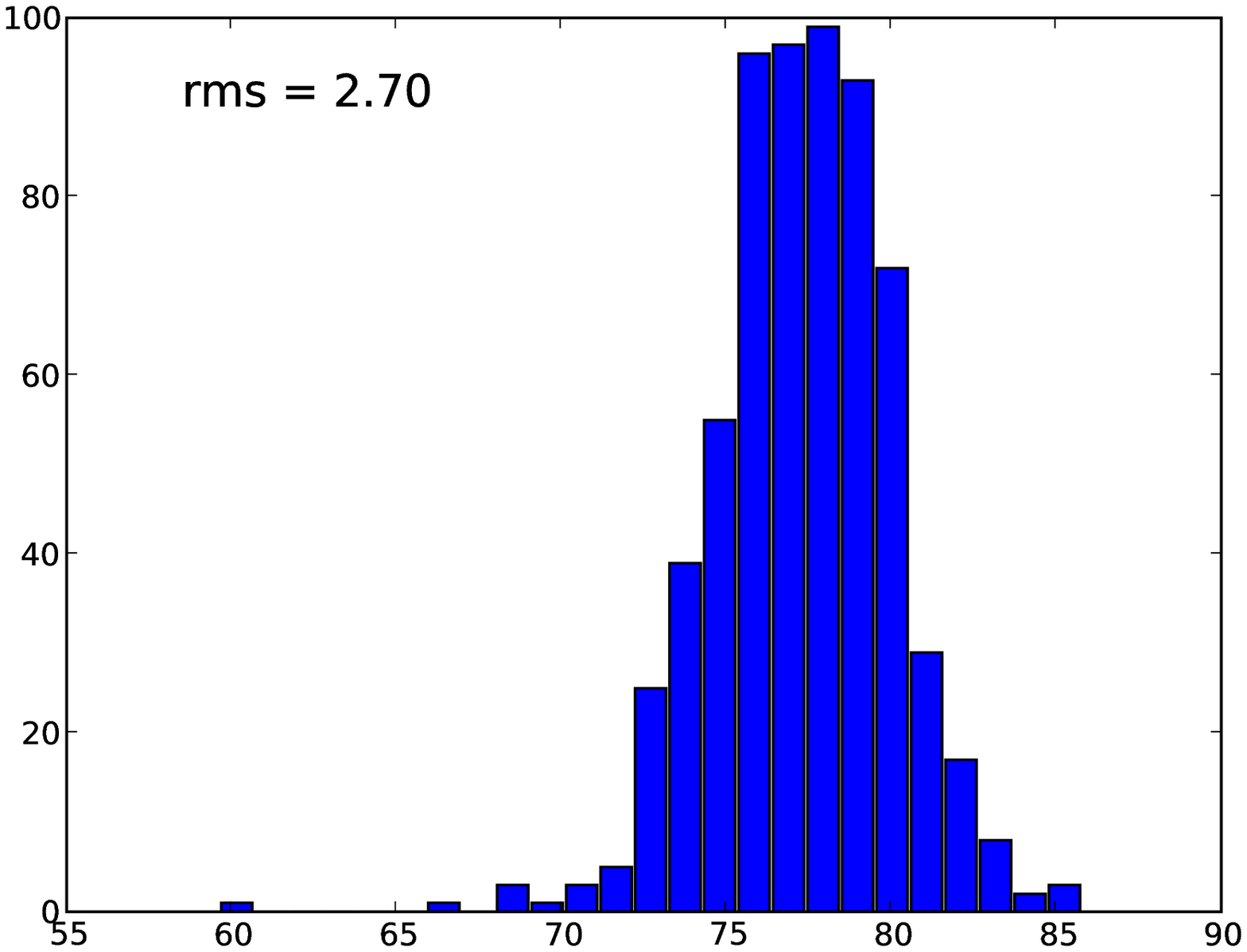}
\vskip 1 true cm
\includegraphics[width= 8 cm,angle=270,clip]{cerna_figure4b.epsi}
\caption{a) flux distribution ADCU/pixel/frame  for an LED setting 
            at 40 $\mu$A 
         b) variance $(ADCU)^2$ and flux (ADCU) as a function of 
            the ADC value, for a LED setting at 20 $\mu$A.}
\end{center}
\end{figure}

\newpage
\begin{figure}[!htb]
\includegraphics[width=0.90\textwidth,clip]{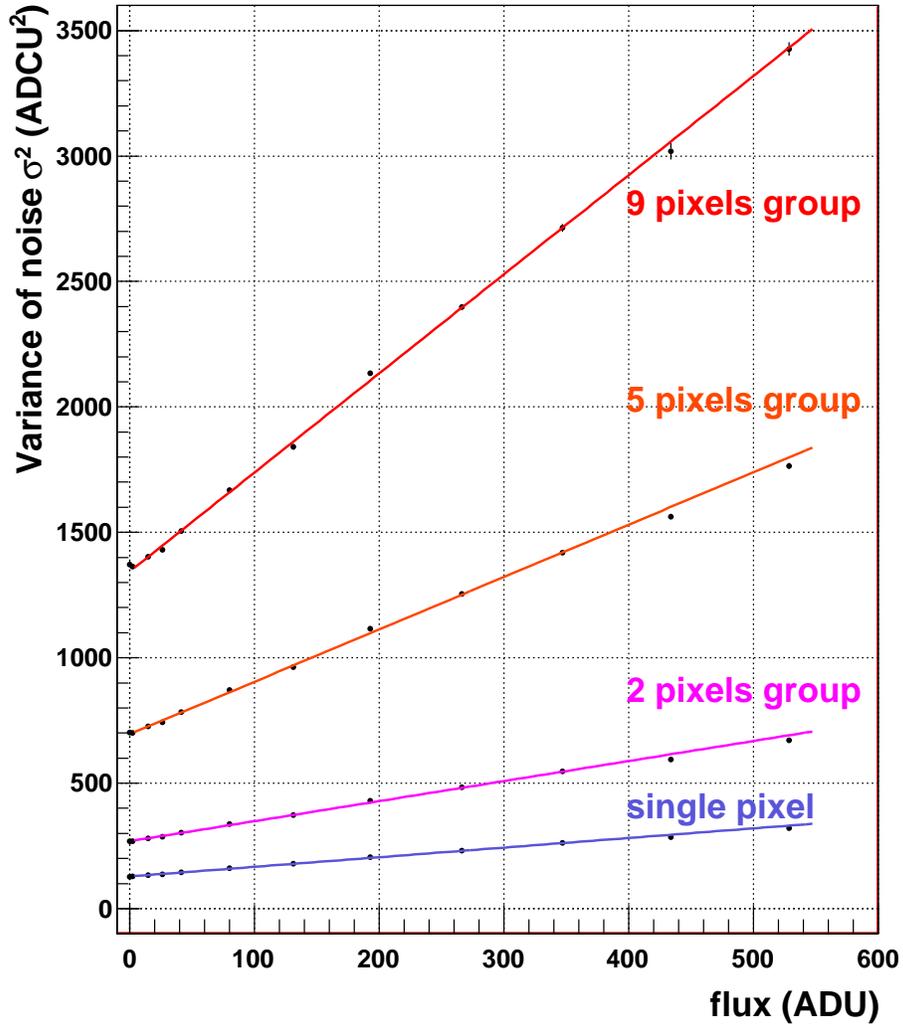}
\caption{noise(without reference subtraction)/Flux for single pixels and groups of 2,5 and 9 pixels}
\end{figure}

\end{document}